\documentclass[11pt,a4paper]{article}

\usepackage{epsfig,amsmath,amssymb,cite}

\begin{document}

\title{Strong deflection gravitational lensing}

\author{Ernesto F. Eiroa$^{1, 2}$\thanks{e-mail: eiroa@iafe.uba.ar}\\
{\small $^1$ Instituto de Astronom\'{\i}a y F\'{\i}sica del Espacio, C.C. 67, 
Suc. 28, 1428, Buenos Aires, Argentina}\\
{\small $^2$ Departamento de F\'{\i}sica, Facultad de Ciencias Exactas y 
Naturales,} \\ 
{\small Universidad de Buenos Aires, Ciudad Universitaria Pab. I, 1428, 
Buenos Aires, Argentina}}

\date{}

\maketitle

\begin{abstract}

In this work are reviewed several aspects of gravitational lensing produced by astrophysical bodies that strongly curve the spacetime in their vicinity. When an object with a photon sphere (e.g. a black hole) is interposed between a source and an observer, in addition to primary and secondary images due to small deflections of the light rays, there are two infinite sets of images called relativistic, produced by light passing close to the photon sphere, which undergoes a large deflection. The positions and magnifications of relativistic images can be obtained numerically or analytically, in the latter case by an approximate method called strong deflection limit. This method is applied to the study of lensing by non-rotating black holes and also by rotating ones. For the latter, is also of interest to study the deformation of the shadow cast by the black hole, which depends on the rotation parameter.\\

\noindent 
PACS number(s): 98.62.Sb, 97.60.Lf, 98.62.Js

\end{abstract}

\section{Introduction}

The presence of a massive body produces the deflection of light passing close to the object, accordingly to the theory of General Relativity. In 1936, A. Einstein \cite{einstein} pointed out that the deflection of the light of a background star by another star produces two images which can have a great magnification in the case of high alignment of the observer, the lens, and the source. He also found that the angular separation of the images was too small to be resolved by the optical telescopes available at that time. The discover of quasars in 1963 opened up the possibility of really observing gravitational lensing effects. These objects, situated at cosmological distances, are very bright and have a compact optical emitting region. The magnification can be large and the images are well separated in some particular cases. The first gravitational lens discovered, by Walsh et al. in 1979, was the quasar QSO 0957+561 A,B. The weak deflection theory of gravitational lensing, developed, among others, by Y. G. 
Klimov, S. Liebes, S. Refsdal, R. R. Bourossa, and R. Kantowski, has been successful in explaining the astronomical observations up to now. This theory is based on a first order expansion of the small deflection angle. For a detailed treatment see \cite{schneider}, and references therein.

The theoretical research of gravitational lensing by compact objects with a photon sphere, such as black holes and naked singularities, has received great attention in the last decade, mainly because of the strong evidence about the presence of supermassive black holes at the center of galaxies, including the Milky Way \cite{guillessen}. For these lenses, large deflection angles are possible for photons passing close to the photon sphere. These photons could even make one or more complete turns, in both directions of rotation, around the black hole before eventually reaching an observer. As a consequence, two infinite sequences of images, called relativistic, are formed at each side of the black hole. An useful analytical method for obtaining the positions, magnifications, and time delays of the relativistic images corresponding to black holes as gravitational lenses, is the strong deflection limit, a logarithmic approximation of the deflection angle for light rays passing close to the photon sphere. 
Introduced by Darwin \cite{darwin} for the Schwarzschild geometry, it was rediscovered several times \cite{luminet,ohanian,nemiroff,bozza01}, extended to the Reissner-Nordstr\"om spacetime \cite{eiroto}, and to any spherically symmetric black holes \cite{bozza02}.  Numerical studies of black hole lenses were done too \cite{virbha1,virbha2,virbha3}. Black holes with spherical symmetry coming from alternative theories \cite{ei06,sarkar}, string theory or braneworld cosmologies \cite{bhadra,frolov,maj,ei05a,ei05b,whisker,boehmer,binnun10a,binnun10b} and even naked singularities \cite{virbha98,virbha02} were considered as lenses. Retrolenses were also studied \cite{holtz,depaolis1,eito,depaolis2,bozman04}. Particular interest have received the supermassive Galactic black hole as a possible lens, with the stars in its neighborhood as light sources \cite{bozman04,bozman05,bozman09,binnun11}. For other related works see Refs. \cite{atkinson,claudel,bozza04,bozza08,eirom,sendra}. Kerr black hole lenses were analyzed 
by several authors \cite{vazquez,bozza06,bozza07,bozza08ca,kraniotis,keeton1,keeton2}. Rotating black holes present apparent shapes or shadows with an optical deformation due to the spin \cite{bardeen,chandra,young}, instead of being circles as in the case of non-rotating ones. This subject have been re-examined in the last few years \cite{falcke,devries,bozza07,taka,zakha05a,zakha05b,hioki08,bambi,hioki09,amarilla1}, with the expectation that the direct observation of black holes will be possible in the near future \cite{zakha05a,zakha05b,zakha11}, so the study of the shadows will be useful for measuring the properties of astrophysical black holes. Optical properties of rotating braneworld black holes were studied in Refs. \cite{schee,amarilla2}. Other interesting topics are discussed in the recent review articles \cite{perlick,bozza10}.

In this paper, several aspects of strong deflection gravitational lensing due to black holes or naked singularities are reviewed. The first part of this work corresponds to lenses with spherical symmetry and the second part to rotating lenses. In the latter case, particular attention is given to the deformation of the shadow due to the spin. Finally, a brief discussion about the observational prospects is given and a summary is made. Units such as $G=c=1$ are adopted.

\section{Spherically symmetric lenses}

The lensing scenario to be analyzed here consists of a non-rotating object with a photon sphere, a point source of light and an observer. The geometry is supposed asymptotically flat, with the source and the observer both situated far from the object, in the flat region\footnote{For a source located at an arbitrary distance see Ref. \cite{bozza07}.}.

\subsection{Deflection angle}

\begin{figure}[t!]
\centering
\includegraphics[width=0.9\textwidth]{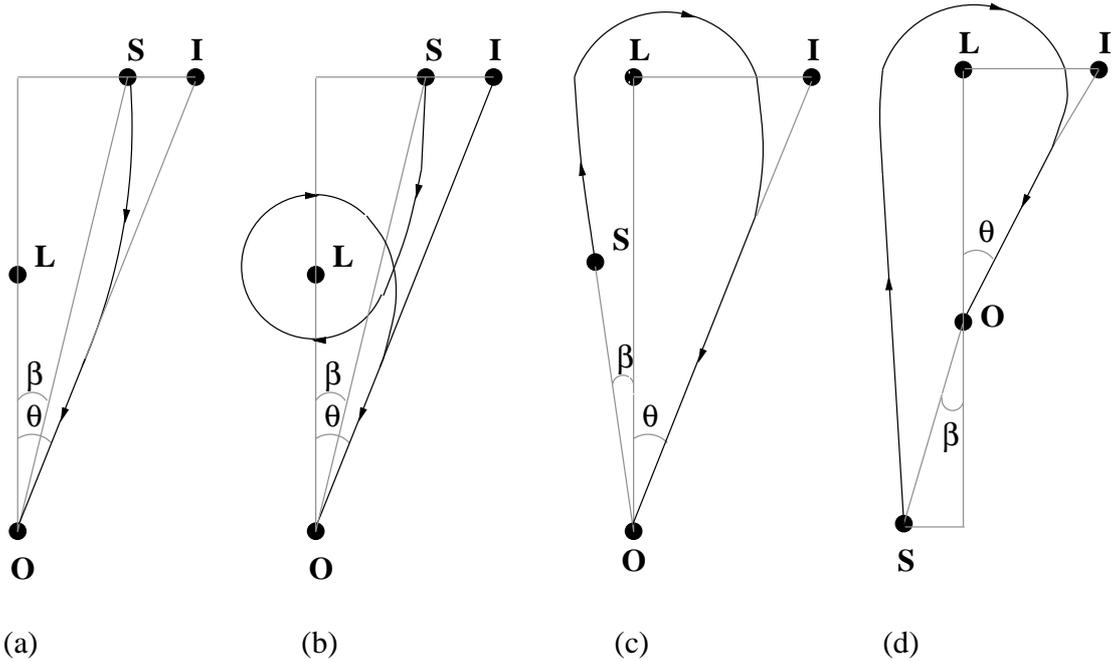}
\caption{Schematic lensing diagrams, where the point light source (S), the deflector (L), the observer (O), and the image (I) are shown. The angular diameter distances are $d_{\mathrm{OS}}$ (observer-source), $d_{\mathrm{OL}}$ (observer-lens), and $d_{\mathrm{LS}}$ (lens-source); $\beta $ and $\theta $ are respectively the angular source and image positions. From left to right, the first two diagrams (a) and (b), correspond to standard lensing, with the lens between the observer and the source, and the other two (c) and (d), to retrolensing.}
\label{ld}
\end{figure}

For a general spherically symmetric and asymptotically flat metric, having in Schwarzchild coordinates the form
\begin{equation}
ds^{2}=-f(r)dt^{2}+g(r)dr^{2}+h(r)(d\theta ^{2}+\sin ^2 \theta d\varphi ^2),
\end{equation}
the deflection angle $\alpha $ is given by \cite{weinberg,virbha98}
\begin{equation}
\alpha (r_{0})=-\pi +\int_{r_{0}}^{\infty }
2\left[ \frac {g(r)}{h(r)}\right] ^{1/2}
\left[ \frac {h(r)f(r_{0})}{h(r_{0})f(r)}-1\right] ^{-1/2}dr,
\label{alpha}
\end{equation}
where $r_{0}$ is the closest approach distance. The photon sphere radius $r_{ps}$, corresponding to the unstable circular photon orbits, is the largest positive solution \cite{atkinson,claudel} of the equation:
\begin{equation}
\frac{f'(r)}{f(r)}=\frac{h'(r)}{h(r)};
\label{ps}
\end{equation}
for Schwarzschild geometry $r_{ps}=3M$.
 
There are two cases where the deflection angle $\alpha$ can be approximated by simple expressions:
\begin{itemize}
\item Weak deflection limit: if $r_{0}\gg r_{ps}$, the deflection angle can be approximated by the first non null term of the Taylor expansion in $1/r_{0}$ (e.g. Ref. \cite{schneider}). This limit is suitable for the analytic calculations corresponding to the primary and secondary images, formed by photons passing at large distances from the deflector.
\item Strong deflection limit: the deflection angle $\alpha (r_{0})$ diverges when $r_{0}\rightarrow r_{ps}$. For $0<r_{0}/r_{ps}-1\ll 1$, it can be approximated by a logarithmic function \cite{bozza02}:
\begin{equation}
\alpha(r_{0})= -a_{1}\ln \left( \frac{r_{0}}{r_{ps}}-1 \right) +a_{2}+ \mathrm{O} \left( \frac{r_{0}}{r_{ps}}-1 \right),
\label{sdl1}
\end{equation}
where $a_{1}$ and $a_{2}$ are constants that depend on the specific form of the metric functions (see Appendix); for the Schwarzschild black hole the values can be calculated exactly \cite{bozza01}: $a_{1}=2$ and $a_{2}=\ln [144(7-4\sqrt{3})]-\pi$. This limit is useful for the analytic computations corresponding to the two sets of strong deflection (or relativistic) images, formed by photons passing close to the photon sphere. The strong field limit provides an excellent approximation for the deflection angle (see discussion in Refs. \cite{bozza01,bozza04}).
\end{itemize}
In Fig. \ref{ld}, different possible lensing situations are shown. The optical axis is defined as the line joining the observer and the lens. The angular positions of the source and the images, seen from the observer, are $\beta $ (taken positive) and $\theta $, respectively. The observer-source ($d_{\mathrm{OS}}$), observer-lens ($d_{\mathrm{OL}}$) and the lens-source ($d_{\mathrm{LS}}$) distances are measured along the optical axis and they are taken much larger than the horizon radius. From left to right, the first two plots (a) and (b), correspond standard lensing, i.e. the deflector between the source and the observer, while the other two (c) and (d), represent retrolensing scenarios \cite{holtz,eito}. The first diagram (a) corresponds to photons with $r_0\gg r_{ps}$, which form the primary and secondary images (only one of them is shown). The second plot (b) shows a light ray which makes a complete turn around the lens before reaching the observer; in this way the first relativistic image is formed (
there is another one due to light rays passing by the other side of the lens, not shown). Other relativistic images are obtained with photons making more turns around the deflector, not shown in the diagram. The third (c) and four (d) plots show one retrolensing image produced by photons with a deflection angle close to $\pi $ (light rays passing by the other side of the lens form another image, not shown). More retrolensing images are formed by photons that make complete turns around the lens, with a deflection angle close to an odd multiple of $\pi$, not shown in the diagrams.

\subsection{Primary and secondary images}

In the case of standard lensing with a very small source angle $\beta $, the primary and secondary images are present, for which the deflection angle $\alpha $ is very small. Then, the angular positions of the images are also small.

\subsubsection*{Image positions}

It is not difficult to see from the lens diagram, that the lens equation for small lensing angles takes the form \cite{schneider}:
\begin{equation}
\beta =\theta -\frac{d_{\mathrm{LS}}}{d_{\mathrm{OS}}}\alpha.
\end{equation}
The first step is to calculate the deflection angle $\alpha $ in the weak deflection limit, by Taylor expanding the integrand of Eq. (\ref{alpha}) and performing the integral. To first order in $1/r_0$, one finds that $\alpha \approx \chi /r_0$, where $\chi$ is a constant which depends on the metric considered; for Schwarzschild geometry $\chi = 4M$. Using that $r_0=d_{\mathrm{OL}}\theta$ and inverting the lens equation, the angular position of the images are:
\begin{equation}
\theta_{p,s}=\frac{1}{2}\left(\beta \pm\sqrt{\beta ^2+4 \theta_E^2}\right),
\label{thps}
\end{equation}
with
\begin{equation}
\theta_E=\sqrt{\frac{\chi d_{\mathrm{LS}}}{d_{\mathrm{OS}}d_{\mathrm{OL}}}}
\end{equation}
the Einstein radius. For perfect alignment, instead of two point images, an Einstein ring with angular radius $\theta_E$ is obtained.

\subsubsection*{Magnifications}

Gravitational lensing conserves surface brightness (e.g. Ref. \cite{schneider}), then the magnification is the quotient of the solid angles subtended by the image and the source
\begin{equation}
\mu =\left| \frac{\sin \beta }{\sin \theta }
\frac{d\beta }{d\theta }\right|^{-1}\approx 
\left| \frac{\beta }{\theta } \frac{d\beta }{d\theta }\right|^{-1},
\label{mu}
\end{equation}
which, using Eq. (\ref{thps}), gives for each image:
\begin{equation}
\mu_{p,s}= \frac{\beta ^2+2}{2\beta \sqrt{\beta^2+4\theta_E^2}}\pm \frac{1}{2},
\end{equation}
and the total magnification is
\begin{equation}
\mu = \frac{\beta ^2+2\theta_E^2}{\beta \sqrt{\beta^2+4\theta_E^2}}.
\end{equation}
The magnifications are proportional to $1/ \beta $, so the magnifications grow as the objects are more aligned.

\subsubsection*{Time delay}

Photons that form the primary and secondary images follow different paths and the time delay between them is \cite{schneider}
\begin{equation}
\Delta t_{p,s}=\chi (1+z_{d})\left( \frac{-\beta \sqrt{\beta^{2}+4\theta_{E}^{2}}}{2\theta_{E}^{2}}+\ln \left| \frac{\beta -\sqrt{\beta^{2}+4\theta_{E}^{2}}}{\beta +\sqrt{\beta^{2}+4\theta_{E}^{2}}}\right|\right),
\end{equation}
where $z_{\rm{d}}$ is the redshift of the deflector. When $\beta =0$ an Einstein ring is obtained, so there is no time delay for perfect alignment. Large time delays can be obtained if $\beta /\theta_{\rm{E}}\gg 1$, but the magnification of the primary image is close to one and the secondary image is very faint in this case. The time delay between the images can be a useful observable quantity in the case of transient sources. The optimal situation for a variable source is when $\beta /\theta_{\rm{E}}$ is small enough to have large magnifications of both images, but not too close to zero, so the time delay can be longer than the typical time scale of the source.

\subsection{Relativistic images}

Light rays passing close to the photon sphere will suffer a large deflection angle. As a consequence, the relativistic images appear, produced by photons making one or more turns around the deflector. The lensing effects are more important when the objects are highly aligned, so here is studied in detail only this configuration, which is the simplest and more interesting one\footnote{For a more general treatment, without the small angle approximation, see Ref. \cite{bozman04}.}. In this case, the angles $\beta $ and $\theta $ are small, and $\alpha $ is close to a multiple of $\pi $. The strong deflection limit is adopted for the (approximate) analytic calculations, in which the deflection angle is given by Eq. (\ref{sdl1}), with the coefficients $a_1$ and $a_2$ depending on the specific form of the metric. 

\subsubsection*{Image positions}

The impact parameter $b$ is more easily connected to the lensing parameters than the closest approach distance $r_0$. They are related by the equation \cite{weinberg,virbha98}
\begin{equation}
b=\left[\frac{h(r_0)}{f(r_0)} \right]^{1/2}.
\label{b1}
\end{equation}
By Taylor expanding Eq. (\ref{b1}) and using Eq. (\ref{ps}) to obtain that $b'(r_{ps})=0$, it is straightforward to see that
\begin{equation}
b-b_{ps}\approx k (r_{0}-r_{ps})^2,
\label{b2}
\end{equation}
with $b_{ps}=\left[h(r_{ps})/f(r_{ps}) \right]^{1/2}$ and $k=b''(r_{ps})/2$ (for Schwarzschild lens,  $b_{ps}=3\sqrt{3}M$, $k=\sqrt{3}/(2M)$). Then, by replacing Eq. (\ref{b2}) in Eq. (\ref{sdl1}), one obtains
\begin{equation}
\alpha (\theta )=-c_{1}\ln \left( \frac{b}{b_{ps}}-1 \right) 
+c_{2}+\mathrm{O}\left( \frac{b}{b_{ps}}-1 \right),
\label{sdl2} 
\end{equation}
where 
\begin{equation}
c_{1}=\frac{a_{1}}{2}
\end{equation}
and
\begin{equation}
c_{2}=a_{2}+\frac{a_{1}}{2} \ln \left( \frac{k r_{ps}^2}{b_{ps}}\right).
\end{equation}
For Schwarzschild metric, the values are \cite{bozza02}: $c_1=1$ and $c_2=\ln [216(7-4\sqrt{3})]-\pi $. Photons having an impact parameter $b$ slightly larger than the critical value $b_{ps}$ will spiral out, eventually reaching an observer after one or more turns around the black hole, and those with $b$ smaller than $b_{ps}$ will spiral into the black hole, not reaching any observer outside the photon sphere. From the lens geometry, it can be seen that
\begin{equation}
b=d_{\mathrm{OL}}\sin \theta \approx d_{\mathrm{OL}}\theta ;
\label{b3}
\end{equation}
so, using Eqs. (\ref{b3}) and (\ref{sdl3}), the strong deflection limit takes the form
\begin{equation}
\alpha (\theta )=-c_{1}\ln \left( \frac{\theta }{\theta _{ps}}-1 \right) 
+c_{2}+\mathrm{O}\left( \frac{\theta }{\theta _{ps}}-1 \right),
\label{sdl3} 
\end{equation}
with  $\theta _{ps}=b_{ps}/d_{\mathrm{OL}}$. The lens equation that relates the lensing angles with the distances can be written in the form \cite{virbha1}\footnote{For a throughout discussion about possible lens equations see Ref. \cite{bozza08}.}:
\begin{equation}
\tan \beta =\tan \theta -c_{3}\left[ \tan (\alpha -\theta ) + \tan \theta \right] ,
\label{le1}
\end{equation}
where $c_{3}=d_{\mathrm{LS}}/d_{\mathrm{OS}}$ \cite{virbha1} for standard lensing (Fig. \ref{ld} (b)) and $c_{3}=d_{\mathrm{OS}}/d_{\mathrm{OL}}$ or $c_{3}=d_{\mathrm{OS}}/d_{\mathrm{LS}}$ for retrolensing (Fig. \ref{ld} (c) and (d), respectively) \cite{eito}. A non-negative angle $\beta $ can be taken without losing generality. When $\beta \neq 0$  two infinite sets of point relativistic images are obtained. \cite{darwin,virbha1}. For high alignment, the deflection angle can be approximated by
\begin{equation}
\alpha =\pm m\pi \pm\Delta \alpha _{m},
\end{equation}
with $m\in \mathbb{N}$ and $0<\Delta \alpha _{m}\ll 1$; $m$ is called the winding number, it is even for standard lensing and odd for retrolensing. The plus and minus signs correspond to light rays passing by different sides of the black hole. Replacing $\alpha $ in Eq. (\ref{le1}), and using that $\beta$ and $\theta $ are small for high alignment, the lens equation adopts the form:
\begin{equation}
\beta =\theta _{m} \mp c_{3}\Delta \alpha _{m}.
\label{le2}
\end{equation}
Inverting Eq. (\ref{sdl3}) to obtain $\theta (\alpha )$
\begin{equation}
\theta =\theta _{ps}\left[ 1+e^{(c_{2}-\alpha )/c_{1}}\right] ,
\end{equation}
and making a first order Taylor expansion around $\alpha =m\pi $, the angular 
position of the $m$-th image is given by
\begin{equation}
\theta _{m}=\theta ^{0}_{m}\mp \zeta _{m}\Delta \alpha _{m},
\label{tm}
\end{equation}
with
\begin{equation}
\theta ^{0}_{m}=\theta _{ps}\left[ 1+e^{(c_{2}-m\pi )/c_{1}}
 \right] ,
\end{equation}
and
\begin{equation}
\zeta _{m}=\frac{\theta _{ps}}{c_{1}}e^{(c_{2}-m\pi )/c_{1}}.
\end{equation}
From Eq. (\ref{le2}), one has that $\Delta \alpha _{m}=\pm (\theta _{m}-\beta )/c_{3}$, and replacing it in Eq. (\ref{tm}) leads to
\begin{equation}
\theta _{m}=\theta ^{0}_{m}\mp \frac{\zeta _{m}}{c_{3}}(\theta _{m}-\beta ),
\end{equation}
which can be written in the form
\begin{equation}
\theta _{m}=\left( 1 \pm \frac{\zeta _{m}}{c_{3}}\right) ^{-1}\left( 
\theta ^{0}_{m} \pm \frac{\zeta _{m}}{c_{3}}\beta \right) .
\end{equation}
Finally, using that $0<\zeta _{m}/c_{3}\ll 1$ and keeping only the first order term in $\zeta _{m}/c_{3}$, the angular positions of the images are given by
\begin{equation}
\theta _{m}=\pm \theta ^{0}_{m}+\frac{\zeta _{m}}{c_{3}}\left( \beta \mp \theta ^{0}_{m}\right) .
\label{thri}
\end{equation}
For perfect alignment an infinite sequence of Einstein rings with angular radius
\begin{equation}
\theta ^{E}_{m}=\left( 1-\frac{\zeta _{m}}{c_{3}}\right) \theta ^{0}_{m}
\end{equation}
is obtained instead of the two sequences of point images.

\subsubsection*{Magnifications}

As gravitational lensing conserves surface brightness (e.g. Ref. \cite{schneider}), the magnification is the quotient of the solid angles subtended by the $m$-th image and the source
\begin{equation}
\mu _{m}=\left| \frac{\sin \beta }{\sin \theta _{m}}
\frac{d\beta }{d\theta _{m}}\right|^{-1}\approx 
\left| \frac{\beta }{\theta _{m}} \frac{d\beta }{d\theta _{m}}\right|^{-1},
\label{mu1}
\end{equation}
and using Eq. (\ref{thri}) it gives
\begin{equation}
\mu _{m}=\frac{1}{\beta}\left[ \theta ^{0}_{m}+
\frac {\zeta _{m}}{c_{3}}(\beta - \theta ^{0}_{m})\right]
\frac {\zeta _{m}}{c_{3}},
\label{mu2}
\end{equation}
so, to first order in $\zeta _{n}/c_{3}$, one has
\begin{equation}
\mu _{m}=\frac{1}{\beta}\frac{\theta ^{0}_{m}\zeta _{m}}{c_{3}},
\label{mu3}
\end{equation}
for both sets of images. The first relativistic image is the brightest one, and the magnifications decrease exponentially with $m$. The total magnification for a point source is obtained by adding the magnifications of  both sets of relativistic images
\begin{equation}
\mu =\frac {1}{\beta}\frac {8}{d_{\mathrm{OL}}^{2}c_{1}c_{3}}
\frac{e^{(c_{2}+\delta )/c_{1}}\left[ 1+e^{(c_{2}+\delta )/c_{1}}+e^{2\pi /c_{1}}
\right] }{e^{4\pi /c_{1}}-1},
\label{mu4}
\end{equation}
with $\delta =0$ for standard lensing and $\delta =\pi $ for retrolensing. The magnifications of the strong deflection images are larger for 
retrolensing. In both cases, the magnifications are proportional to 
$1/d_{ol}^{2}$, which is a very small factor. Then, the relativistic images 
are very faint, unless $\beta $ has values close to zero, i.e. nearly perfect 
alignment. For $\beta =0$, the amplification becomes infinite, and the point 
source approximation breaks down, so an extended source analysis is necessary.

\subsubsection*{Extended source}

When the source is extended, the magnification of the images can be obtained by integration over its luminosity profile:
\begin{equation}
\mu =\frac{\int\!\!\int_{S}\mathcal{I}\tilde{\mu }dS}{\int\!\!\int_{S}
\mathcal{I}dS}, 
\label{ext1}
\end{equation}
where $\mathcal{I}$ is the surface intensity distribution of the source and
$\tilde{\mu }$ is the magnification corresponding to each point of the source. If the source is an uniform disc $D(\beta _{c},\beta _{s})$, with angular radius $\beta _{s}$ and centered in $\beta _{c}$ (taken positive), 
Eq. (\ref{ext1}) can be written in the form
\begin{equation}
\mu =\frac{\int\!\!\int_{D(\beta _{c},\beta _{s})}\tilde{\mu } dS}
{\pi \beta _{s}^{2}}.
\label{ext2}
\end{equation}
Then, using Eq. (\ref{mu2}), the magnification of the relativistic $m$-th 
image (with $m$ even for standard lensing and odd for retrolensing) when the source is an uniform disc is given by \cite{bozza01}
\begin{equation}
\mu _{m}=\frac{I}{\pi \beta _{s}^{2}}\frac {\theta ^{0}_{m}\zeta _{m}}{c_{3}},
\label{ext3}
\end{equation}
where
\begin{equation}
I=2\left[ (\beta _{s}+\beta _{c})
E\left(\frac{2\sqrt{\beta _{s}\beta _{c}}}{\beta _{s}+\beta _{c}}\right)
+(\beta _{s}-\beta _{c})K\left(\frac{2\sqrt{\beta _{s}\beta _{c}}}
{\beta _{s}+\beta _{c}}\right) \right] ,
\label{ext4}
\end{equation}
with $K(k)=\int _{0}^{\pi/2}(1-k^{2}\sin ^{2}\phi )^{-1/2}d\phi $ and  $E(k)=\int_{0}^{\pi /2}\left( 1-k^2\sin ^{2}\phi \right) ^{1/2}d\phi $ the complete elliptic integrals of the the first and the second kind, respectively. The total magnification for an uniform disc is
\begin{equation}
\mu =\frac{I}{\pi \beta _{s}^{2}}\frac {8}{d_{\mathrm{OL}}^{2}c_{1}c_{3}}
\frac{e^{(c_{2}+\delta )/c_{1}}\left[ 1+e^{(c_{2}+\delta )/c_{1}}+e^{2\pi /c_{1}} \right] }{e^{4\pi /c_{1}}-1},
\label{ext5}
\end{equation}
with $\delta =0$ for standard lensing and $\delta =\pi $ for retrolensing. These expressions always give finite magnifications, even in the case of 
complete alignment.

\subsubsection*{Time delays}

As pointed out above, the time delay between images is of particular interest in the case of transient sources. For the standard lensing configuration, the time delay between the relativistic images \cite{bozza04}, resulting from the different paths followed by the photons that form them, at the same side of the lens with winding numbers $m$ and $n$, is given  by\footnote{The expressions from Ref. \cite{bozza04} have been rewritten here, by adding the cosmological factor $1+z_{d}$ and expanding them to first order in the source position angle, which is measured from the observer instead of from the source, as done in Ref. \cite{eirom} for Schwarzschild geometry.}
\begin{eqnarray}
\Delta t^{\rm{o}}_{n,m}&=&b_{ps}(1+z_{d})\left[ 2\pi (n-m)+\left( e^{(c_2-2m\pi)/(2c_1)}-e^{(c_2-2n\pi)/(2c_1)}\right) \kappa \right. \nonumber \\  
&& \left. \pm \frac{d_\mathrm{OS}}{d_\mathrm{LS}}\left( e^{(c_2-2m\pi)/(2c_1)}-e^{(c_2-2n\pi)/(2c_1)}\right) \frac{\kappa }{2c_1} \beta )\right],
\label{rtd1}
\end{eqnarray}
where $\kappa =2\sqrt{g(r_{ps})}/\sqrt{kb_{ps}f(r_{ps})}$, and the plus (minus) sign if both images are on the same (opposite) side of the source . For the images at the opposite side of the lens:
\begin{eqnarray}
\Delta t^{\rm{o}}_{n,m}&=& b_{ps}(1+z_{d})\left\{ 2\pi (n-m) +\left( e^{(c_2-2m\pi)/(2c_1)}-e^{(c_2-2n\pi)/(2c_1)}\right) \kappa \right. \nonumber \\  
&& \left. + \frac{d_\mathrm{OS}}{d_\mathrm{LS}}\left[ \left( e^{(c_2-2m\pi)/(2c_1)}+e^{(c_2-2n\pi)/(2c_1)}\right) \frac{\kappa }{2c_1}-2
\right] \beta \right\},
\label{rtd2}
\end{eqnarray}
where the image with winding number $n$ is on the same side of the source and the one with $m$ on the opposite side. The first term in Eqs. (\ref{rtd1}) and (\ref{rtd2}) is by large the most important one. The time delays between the relativistic images are longer than the time delay between the primary and the secondary images.

\section{Rotating lenses}

In the case of rotating black holes, photons follow complicated paths in the vicinity of these compact objects. They do not move in a single plane, so the problem is more complex than in the non-rotating case. The Hamilton-Jacobi equation determines the null geodesics for a given geometry. In the case of Kerr metric the problem can be separated by using Carter constants (e.g. Ref. \cite{chandra}). In other spacetimes this separation can be also done. The caustic structure is also more complex than in the non-rotating case \cite{bozza08ca}. The analytical treatment of rotating black hole lenses is much difficult than non-rotating ones. The strong deflection limit was extended to Kerr black hole lenses to obtain the positions and magnifications of the relativistic images, to second order in the rotation parameter $a$. For a detailed treatment, see \cite{bozza06} and references therein.

\subsubsection*{Shadows}

\begin{figure}[t!]
\centering
\includegraphics[width=0.3\textwidth]{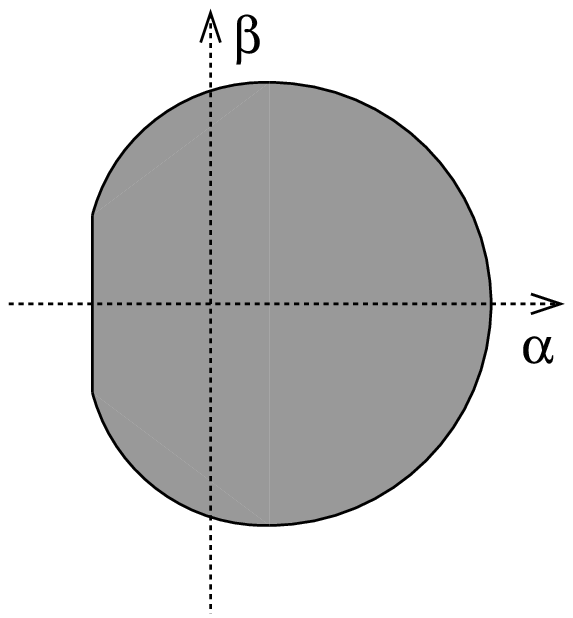}
\hspace{0.1\textwidth}
\includegraphics[width=0.3\textwidth]{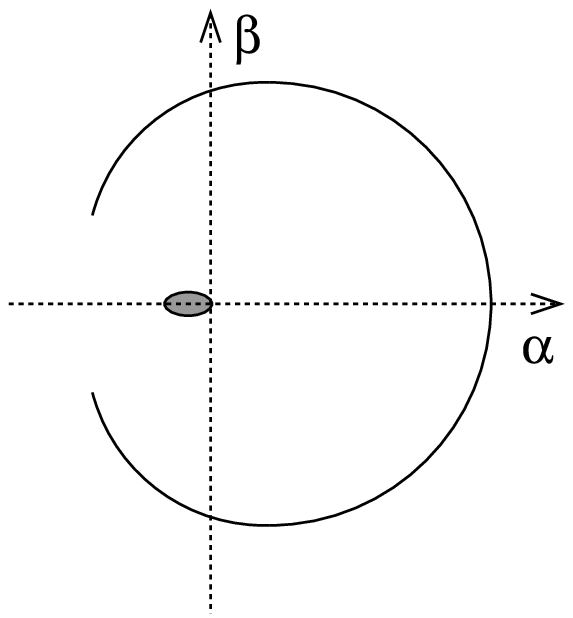}
\caption{Shadows of a rotating black hole (left) and a rotating naked singularity (right).}
\label{shadow}
\end{figure}

When a black hole is in front of a luminous background, the light reaches the observer after being deflected by the black hole gravitational field; but part of the photons emitted by the source, those with small impact parameters, end up falling into the black hole, not reaching the observer, and producing a  completely dark zone called the shadow. The apparent shape of a black hole is thus defined by the boundary of the shadow. For rotating black holes, prograde photons can get closer to the black hole than retrograde ones. The shadows present a deformation that grows with the rotation parameter, instead of being circles as in the non-rotating case. A schematic example is shown in Fig. \ref{shadow} (left). The deformation also grows with the inclination angle $\theta _i $ of the observer with respect to the rotation axis of the black hole. For a polar observer ($\theta_{i}=0$ or $\theta_{i}=\pi$) there is no deformation, while for one in the equatorial plane ($\theta_{i}=\pi/2$) the gravitational effects on 
the shadow are larger. If there are other parameters associated to the black hole, e.g. the charge, it will depend on them too, in a complex way. The contour of the shadow for a given geometry can be obtained by following the procedure shown in Ref. \cite{chandra} for Kerr metric. The first step is to separate the Hamilton-Jacobi action for the null geodesics into a radial and an angular part. The null geodesics are then parametrized in terms of the conserved quantities $\xi =L_z/E$ and $\eta =\mathcal{Q}/E^2$ (with $E$ the energy, $L_z$ the axial component of the angular momentum, and $\mathcal{Q}$ the Carter constant). The quantities $\xi $ and $\eta $  can be related with the celestial coordinates $\alpha $ and $\beta$, which are the apparent perpendicular distances of the image, as seen from the axis of symmetry and from its projection on the equatorial plane, respectively (see e.g. \cite{vazquez}). The boundary of the shadow is determined by the geodesics of photons with parameters $\xi _{sph}$ and $\
eta _{sph}$, corresponding to the unstable spherical orbits around the deflector. This method was applied to Kerr-Newman \cite{devries} and to from alternative theories black holes \cite{hioki08,amarilla1,amarilla2}.

For the characterization of the form of the shadow, are useful the observables defined in Ref. \cite{hioki09}: the radius $R_{s}$ and the distortion parameter $\delta _{s}$. The quantity $R_s$ is the radius of a reference circle passing by three points: the top position  of the shadow, the bottom position  of the shadow, and the point corresponding to the unstable retrograde circular orbit when seen from an observer on the equatorial plane. The distortion parameter is defined by $D/R_s$, where $D$ is the difference between the endpoints of the circle and of the shadow, both of them at the side corresponding to the prograde circular orbit. The radius $R_s$ basically gives the approximate size of the shadow, while $\delta_s$ measures its deformation with respect to the reference circle (see \cite{hioki09} for more details). 

Another related topic is the apparent shape of a rotating naked singularity (if these objects exist in nature); in this case an open dark arc and a deformed dark disc appear (see e.g. \cite{devries,hioki09}), as shown in Fig. \ref{shadow} (right). The unstable spherical photon orbits with a positive radius result in an open dark arc instead of a closed curve; the photons near both sides of the arc can reach the observer due to the nonexistence of the horizon. The unstable spherical photon orbits with a negative radius constructs the small dark disc: the observer will never see the light rays from such directions because they escape into the other infinity by passing through the inside of a singular ring. When the observer is on the equatorial plane ($\theta _i =\pi /2$), the same arc exists but the dark disc disappears, because the light rays in the direction of negative radius will always hit on the ring singularity. Those null geodesics result a straight line. From the arc, two observables can be defined \
cite{hioki09}: the radius $R_a$ of a circumference that approximates the dark arc, and the central angle $\varphi _a$, determined by the angle subtended by the arc, seen from the center of the circumference used to define $R_a$.

\section{Observational prospects}

Forthcoming instruments are expected to look into the region close to Sgr A*, the supermassive black hole at the Galactic center, with mass $4.3 \times 10^{6}M_{\odot}$ and at a distance of $8.3$ kpc \cite{guillessen}. The apparent angular radius of this supermassive black hole is about  $27 \, \mu \mathrm{as}$, so resolutions of the order of $\mu \mathrm{as}$ are required for the observation of the lensing effects considered here. More massive black holes can be found in other galaxies, but the distances are larger too.  In the near future some observational facilities, most of them space-based, will be fully operational, and will be able to measure in the radio and X bands. Two of them are worth mentioning:
\begin{itemize}
\item RADIOASTRON: space-based radio telescope,  launched in July 2011. It will be capable of carrying out measurements with high angular resolution, about $1-10 \, \mu \mathrm{as}$ \cite{zakha05a,zakha05b,zakha11,webradio}.
\item MAXIM project: is a space-based X-ray interferometer with an expected angular resolution of about $0.1 \, \mu \mathrm{as}$ (see \cite{webmaxim} for further details). 
\end{itemize}
The observation of black hole apparent shapes is a major goal in observational astrophysics, since those shadows correspond to a full description of the near horizon region, without any theoretical assumption concerning the underlying theory or astrophysical processes in the black hole surroundings. These instruments will be able to resolve the shadow of the supermassive Galactic black hole, which together with other observations, would serve to obtain its parameters in the near future. Gravitational lensing would be then an useful tool for the study of black holes. More subtle effects, like the comparison with observations of different models of black holes, corresponding to alternative theories of gravity, would require a future second generation of instruments. Other interesting observational aspects are discussed in Refs. \cite{bozza10,zakha11}.

\section{Summary}

The observation of the strong deflection effects will be a very difficult but not impossible task not so far in the future. Different topics of strong deflection gravitational lensing produced by a black hole or a naked singularity were reviewed in this paper. The weak and the strong deflection limits were used for the analytical computation of the positions, magnifications and time delays of the primary, secondary and relativistic images, for high alignment, which is  the simplest and more interesting case. The relativistic images have an angular position about the size of the angle subtended by the event horizon and they are strongly demagnified, making their detection a challenge for future instruments. The time delay between images for transient sources was also discussed in this work. Other relevant aspect is the deformation of the shadow for rotating lenses, which grows with the spin, and it would give useful information in the near future, when direct imaging of the Galactic supermassive black hole 
will be possible.

\section*{Acknowledgments}

This work was supported by CONICET and Universidad de Buenos Aires. The Author would like to thank the organizers of GAC I for their kind invitation to the meeting.

\section*{Appendix}

In this Appendix, the strong deflection limit coefficients are calculated. The deflection angle can be rewritten in the form $\alpha(r_0)=I(r_0)-\pi $, where $I(r_0)$ is the integral in Eq. (\ref{alpha}). The integral $I(r_0)$ grows as $r_0$ approaches to $r_{ps}$, where it diverges. Following Ref. \cite{bozza02}, it is useful to define a new variable $z=\left[ f(r)-f(r_0)\right] / \left[ 1-f(r_0)\right] $, and the functions
\begin{equation}
R(z,r_0)=\frac{2\sqrt{f(r)g(r)}}{f'(r) h(r)}\left[ 1-f(r_0) \right] \sqrt{h(r_0)}, 
\label{rz}
\end{equation}
\begin{equation}
w(z,r_0)=\frac{1}{\sqrt{f(r_0)-\left[\left(1-f(r_0)\right)z+f(r_0)\right]h(r_0) [h(r)]^{-1}}}, 
\label{fz}
\end{equation}
where $r=f^{-1}[(1-f(r_0))z+f(r_0)]$. By performing a Taylor expansion of the function inside the square root in Eq. (\ref{fz}) one obtains
\begin{equation}
w_0(z,r_0)=\frac{1}{\sqrt{\varphi (r_0) z+\gamma (r_0) z^{2}}}, 
\label{f0}
\end{equation}
where
\begin{equation}
\varphi (r_0)=\frac{1-f(r_0)}{f'(r_0) h(r_0)}\left[ f(r_0) h'(r_0) - f'(r_0) h(r_0)\right],
\label{varphi}
\end{equation}
and
\begin{eqnarray}
\gamma (r_0) &=& \frac{\left[ 1-f(r_0)\right] ^{2}}{2[f'(r_0)]^{3} [h(r_0)]^{2}}\left\{ 2 [f'(r_0)]^{2} h(r_0) h'(r_0) - f(r_0) f''(r_0) h(r_0) h'(r_0) \right. \nonumber \\
&& \left. + f(r_0) f'(r_0) \left[ h(r_0) h''(r_0) -2 [h'(r_0)]^{2}\right] \right\} .
\label{gamma}
\end{eqnarray}
With these definitions, the integral $I(r_0)$ can be separated into two parts
\begin{equation}
I(r_0)=I_D(r_0)+I_R(r_0), 
\label{i0n}
\end{equation}
where 
\begin{equation}
I_D(r_0)=\int^{1}_{0}R(0,r_{ps})w_0(z,r_0)dz, 
\label{id}
\end{equation}
and
\begin{equation}
I_R(r_0)=\int^{1}_{0}[R(z,r_0)w(z,r_0)-R(0,r_{ps})w_0(z,r_0)]dz. 
\label{ir}
\end{equation}
If $r_0 \neq r_{ps}$ we have that $\varphi \neq 0$ and $w_0\sim 1/\sqrt{z}$, so the integral $I_D(r_0)$ converges. Instead, when $r_0= r_{ps}$, by using Eq. (\ref{ps}) one has that $\varphi =0$ and $w_0\sim 1/z$, so $I_D(r_0)$ has a logarithmic divergence. Therefore, $I_D(r_0)$ is the term containing the divergence at $r_0=r_{ps}$ and $I_R(r_0)$ is regular because it has the divergence subtracted. By performing the integral in Eq. (\ref{id}) and making a Taylor expansion, it can be seen that \cite{bozza02}
\begin{equation}
I_D(r_0)=-a_{1}\ln \left( \frac{r_{0}}{r_{ps}}-1 \right) +a_{\mathrm{D}}+ \mathrm{O} \left( \frac{r_{0}}{r_{ps}}-1 \right),
\end{equation}
where
\begin{equation}
a_{1}=\frac{R(0,r_{ps})}{\sqrt{\gamma (r_{ps})}}, 
\label{a1}
\end{equation}
and
\begin{equation}
a_{\mathrm{D}}=\frac{R(0,r_{ps})}{\sqrt{\gamma (r_{ps})}}\ln\frac{2[1-f(r_{ps})]}{r_{ps}f'(r_{ps})}. 
\end{equation}
Finally, by Taylor expanding the regular integral $I_R(r_0)=I_R(r_{ps}) + \mathrm{O} (r_0 / r_{ps}-1)$, and defining $a_{\mathrm{R}}=I_R(r_{ps})$, the strong deflection limit takes the form shown in Eq. (\ref{sdl1}) with $a_1$ given by Eq. (\ref{a1}) and  
\begin{equation}
a_{2}=a_{\mathrm{R}}+a_{\mathrm{D}}-\pi . 
\label{a2}
\end{equation}
In some cases, e.g. in Reissner-Nordstr\"om spacetime, the integral $I_R(r_{ps})$ cannot be calculated exactly and it is necessary a numerical integration or a Taylor expansion in terms of a relevant parameter to obtain $a_R$. For an alternative numerical computation of the strong deflection limit coefficients see Ref. \cite{eiroto}.

\end{document}